\documentclass[runningheads
              ]{llncs}

\usepackage[utf8]{inputenc}
\usepackage{url}
\usepackage{slashbox}
\usepackage{rotating}
\usepackage{array}
\usepackage{hyperref}

\usepackage{color}
\usepackage[normalem]{ulem}
\def\<#1|#2>{{\color{red}\sout{#1}} {\color{blue}\textbf{#2}}}

\newcommand{\threestar}[0]{{$^{\ast\!\ast\!\ast}$\ }}

\graphicspath{{./images/}}

\title{The Open Graph Archive: A Community-Driven Effort%
      \texorpdfstring{\thanks{This work was initiated at Schloss
      Dagstuhl in seminar 11191 on ``Graph Drawing with Algorithm
      Engineering Methods''.}}{}}

\author{Christian~Bachmaier\inst{1}
\and Franz~J.~Brandenburg\inst{1}
\and Philip~Effinger\inst{2}
\and Carsten~Gutwenger\inst{3}
\and Jyrki~Katajainen\inst{4}
\and Karsten~Klein\inst{3}
\and Miro~Spönemann\inst{5}
\and Matthias~Stegmaier\inst{2}
\and Michael~Wybrow\inst{6}}
\authorrunning{C.~Bachmaier et al.}

\institute{
University of Passau, Germany
\and
Eberhard-Karls-Universität Tübingen, Germany
\and
Technische Universität Dortmund, Germany
\and
University of Copenhagen, Denmark
\and
Christian-Albrechts-Universität zu Kiel, Germany
\and
Monash University, Australia
}

\begin{document}

\maketitle

\begin{abstract}
  In order to evaluate, compare, and tune graph algorithms,
  experiments on well designed benchmark sets have to be performed. Together
  with the goal of reproducibility of experimental results, this creates
  a demand for a public archive to gather and store graph instances. Such
  an archive would ideally allow annotation of instances or sets of graphs
  with additional information like graph properties and references to the
  respective experiments and results.  Here we examine the requirements,
  and introduce a new community project with the aim of producing
  an easily accessible library of graphs.  Through successful community
  involvement, it is expected that the archive will contain a representative
  selection of both real-world and generated graph instances, covering
  significant application areas as well as interesting classes of graphs.
\end{abstract}

\section{Introduction}

In its basic form a \emph{graph} is a set of vertices and a set of
edges connecting some of the vertices together. In this paper we use
this term broadly: edges can be directed or
weighted, there can be multiple edges between two vertices, and
vertices and edges can be labeled. A graph can also have some metadata
associated with it, answering who-, when-, how-, \mbox{why-,} and what-type of
questions about its creation and existence. A \emph{database} is an
organized collection of data usually stored in digital form. A
\emph{graphbase}, a term coined by Knuth~\cite{k-sgb-94}, is a database
of graphs and computer programs that generate, analyze, manipulate, and
visualize graphs.
The terms \emph{graph library} and \emph{graph archive} are often used as synonyms for this term.

Our vision is to provide an infrastructure and quality standards for a
public graphbase, named the Open Graph Archive, that is 
accessible to researchers and other interested parties around the
world via the worldwide web. This paper describes the current work undertaken
towards this goal; the paper is also intended to be a
call for participation since this will be a community-driven effort where
most of the content will be provided by users of the system.

Our motives for building this universal graphbase are similar to
Knuth's motives for building the Stanford GraphBase~\cite{k-sgb-94};
we are just working on a larger scale.  First, we want to provide
standard sets of graphs to enable repeatability of experiments.  We
expect that the graphbase would be particularly interesting for
researchers working in the areas of algorithm engineering and graph
drawing.  Second, we want to provide a single point of access for
datasets relevant for people working with graphs.  By annotating
the datasets with their origin and other semantic information, we can
help researchers to find publications relevant for their work.
Third, a graphbase that is accessible worldwide can stimulate
interesting theory development. As pointed out by Knuth~\cite{k-sgb-94},
a graphbase can bridge the gap between theoreticians
and practitioners. Fourth, the programs (and maybe also the datasets)
available in a graphbase, if done well, can have a significant
educational value.

Many existing collections, like the graphs available in the Stanford
GraphBase~\cite{k-sgb-94} and the well-known Rome graphs~\cite{dgttv-ecfgd-97},
are static and only cover a small number of
data sizes, types, and properties that may be relevant for the
users. In order to allow collection and exchange of interesting
graphs, it is important to make the graphbase extendable. The needs of
the community will certainly change over time. Expandability has
been recognized as an important goal by other researchers as well
(see, e.g.,~\cite{bprbd-mmwrt-97,dh-ufsmc-11}), but the
available data collections seem to be relevant to a limited range of
users only. Our goal is to support the use in a wide variety of application
areas.

This paper is organized as follows. Section~\ref{sec:survey} presents
the results from a survey conducted among a small group of potential
users (the participants of the Dagstuhl seminar 11191) on their needs
and requirements for a useful graphbase. Section~\ref{sec:libs}
reviews and compares existing graph collections.
Section~\ref{sec:grapharchive} describes a prototype implementation
developed at the University of Tübingen that shall form the basis for
further development. Section~\ref{sec:conclusions} concludes with a
call for participation encouraging community involvement in this
endeavour.

\section{User Needs and Requirements}
\label{sec:survey}

In order to investigate the relevance of and the requirements for a
universal graphbase we conducted a survey among 30 participants of the
Dagstuhl seminar 11191, coming from the graph drawing and
algorithm engineering communities.  The survey solicited a variety of
open-ended textual responses.  In this section we summarize the most
interesting and commonly recurring feedback.

\paragraph{Describe two most important use cases for a graph archive.}
The most frequent use cases were to search for graphs with specific
properties, and to benchmark and compare algorithms, both mentioned by
37\% of the participants. Further answers were to share datasets~(27\%),
and to replicate experiments and compare results~(23\%).
Since these are fundamental aspects of experimental scientific
processes, we can see that a graphbase would be an important
tool for researchers of graph algorithms.

\paragraph{What services do you expect?} We proposed nine services of which
the survey participants could select those they considered important.
As shown in Table~\ref{tab:survey_services}, support for tags and arbitrary
comments are the most crucial features. When asked for further
important services, a handful of people wanted to know which
publications refer to a specific graph or collection of graphs~(17\%).

\begin{table}[tb]
  \centering
  \begin{tabular}{@{\hspace{2mm}}l@{\hspace{2mm}}|@{\hspace{2mm}}c@{\hspace{2mm}}}\hline
    \emph{Service} & \emph{Percentage} \\\hline\hline
    Add categorization tags                     & 80\% \\
    Add comments, links, or further information & 77\% \\
    Search for specific tags                    & 77\% \\
    Automatic conversion of file formats        & 70\% \\
    Search for specific properties              & 60\% \\
    Add information on how graphs were created  & 60\% \\
    Add images (drawings of the graphs)         & 50\% \\
    Automatic analysis of graph properties      & 47\% \\
    Programmatic web service                    & 23\% \\
    \hline
  \end{tabular}
  \smallskip
  \caption{Result of the multiple-choice question ``Which services do you
  consider as critical for a graph archive?''}
  \label{tab:survey_services}
\end{table}

\paragraph{Which category tags and analysis properties may be useful?}
Participants named 20 different application domains to categorize a
graph or collection of graphs, e.g., biology, social networks, geography,
software engineering. Furthermore, participants named 16 graph properties, most of
which can be determined automatically. The most popular properties were
connectivity~(60\%), including the number of $k$-connected components,
and planarity~(43\%), including the best known crossing number for
non-planar graphs.

\paragraph{Name two file formats you use most.} The most frequently mentioned
formats were GraphML~(43\%) and GML~(33\%). Since a total of 13 different
formats were named, it is evident that a universal graphbase should not
rely on one specific format, but offer support for several formats,
preferably even converting automatically between formats.

\paragraph{Existing archives and collections.}
Responses for existing archives showed that GraphArchive~\cite{ekms-gogds-11}
from the University of Tübingen and the datasets from the DIMACS implementation
challenges~\cite{dimacs} were both known by a handful of people (20\% and
13\%, respectively). These numbers are quite low and might also be biased
towards the archives used by the researchers that participated in the seminar.
They also suggest that there is currently no commonly used and accepted graph archive service.
Regarding graph collections, participants mostly worked with randomly
generated graphs, as well as with the popular Rome~\cite{dgttv-ecfgd-97}
and AT\&T graphs~\cite{dgttv-ddage-00}.



\paragraph{Community contributions.}
Several participants of the survey declared that they would be willing to provide
human resources (students, testing and development time), a hardware
platform, or even money. This reaffirms that there is definite interest and
enthusiasm for such a system, and also that the project should take
advantage of this through involvement of the community.

\paragraph{Technical and service requirements.}  The survey results
and subsequent community discussions indicate that potential users
agree on a core set of important features, as well as a larger list of
desirable functionality. However, several questions regarding the interface, architecture, and content
remain open.  Below we list the most relevant issues that need to be
discussed or dealt with.

\begin{description}
\item[Storage.]
   Graphs must be stored persistently under a unique ID for
   identification and access.
   Should graphs be stored in their original submission format, or
   converted by the system or the user into a unique storage format?
   In file conversions it is important that as much
   information as possible is preserved.

\item[Metadata.] There is a variety of metadata that can
  be stored with a graph, e.g., creator, description of the underlying
  data or the generator, additional keywords, and links to
  corresponding experiments or publications.  Some of this data should
  be defined as mandatory properties, whereas other parts may be
  added as generic text properties. Useful keywords/tags for
  categorization need to be defined.  Some tags could be attributes for
  graphs or collections of graphs, and some could list their structural
  and semantic properties.

\item[Searching.] Based on the survey results and our own
  experience, we assume that a graphbase should allow the user to search
  using both graph properties (number of nodes, etc.) and annotations
  (categories, origin, etc.).

\item[Data analysis.] Automatic analysis of basic graph
  properties must be possible. However, we are not sure if there
  should be a restriction on the computational
  complexity of the analysis or on the size of the analyzed graphs,
  or if users should be allowed to upload that information, e.g.,
  the crossing number of a graph.

\item[Programs.] In addition to datasets, it must be possible to store
  programs like graph generators, analyzers, or visualizers. If the
  graphbase contains randomly generated collections of graphs with
  certain attributes, it would be useful to provide access to the
  programs used for their generation.

 \item[Ownership and copyright.] The ownership of uploaded
   graphs must be clear from the outset. The content should be as
   freely usable as possible with fair attribution to the original
   authors. Contributors will need to take responsibility for their
   submitted graphs and collections of graphs.

 \item[Existing collections.] Existing popular
 	collections should be made identifiable and accessible via the
	system.
\end{description}

\paragraph{Possible extensions.}
Further useful extensions may include the following:
\begin{itemize}
\item Automatic file
   conversion could be provided as an additional service and the programs
   providing these conversions could also be made available.
\item A series of drawings (layouts) for submitted graphs could be
   provided, or even automatic layout on demand, and the
   programs used for drawing the graphs could be made publicly available.
\item Special support for browsing collections of graphs could be
  provided. For this purpose a hierarchical classification system can
  be useful.
\item Structure-based searching could be supported, e.g., find graphs
   containing a clique of a specific size.
\item Versioning of individual graphs as well as the possibility to
   store a series of dynamic graphs could be supported.
\item A web-service API could be provided to allow interrogation of the
   database by computer programs, rather than via a web browser.
\end{itemize}

\section{Related Work}
\label{sec:libs}

In this section we examine the features of existing systems in more depth.
Based on the survey responses, we selected the most relevant
existing archives and checked their capabilities with respect to the
desired features. Not all of these archives are designated graph or even
graph-drawing archives; several are dedicated to either specific
experimental goals (e.g.,~\cite{dimacs}) or to matrices (e.g.,~\cite{dh-ufsmc-11}), and the interfaces are designed accordingly.
Table~\ref{tab:systems} lists the most desired features from the
survey and evaluates existing systems accordingly.


Together with a large number of customized benchmark sets, several widely
used graph collections have become de-facto standards for benchmarking in
graph drawing.
Note that the following list does not lay claim to completeness. Rather,
it is a selection of graph collections commonly used within
the graph-drawing community. Other popular collections are the GD contest
graphs~\cite{gdcontest}, the test suite from GDToolkit~\cite{gdtoolkit}, the
Hachul graphs~\cite{HachulJ07}, and the graphs in the Stanford GraphBase~\cite{k-sgb-94}.

\begin{list}{xxx}{
    \setlength{\leftmargin}{5mm}\setlength{\labelwidth}{5mm}\setlength{\labelsep}{0mm}\setlength{\itemsep}{1ex}}

\item[The \textbf{Rome library}]~\cite{dgttv-ecfgd-97}
  consists of 11528 small undirected connected graphs with 10 to 100
  vertices with a limited variation of structures and properties. They are
  derived from a small set (112 instances) of (outdated) real-world graphs
  from software-engineering and database applications. The original
  collection with 11582 instances contains some corrupted files and
  duplicates.
  The 112 core instances were extended by executing multiple rounds of
  random sequences of five primitive operations including vertex/edge
  removal and insertion.
  After each iteration the graphs were filtered by
  testing their suitability, e.g., by visual inspection of structural
  similarity. The probability of each primitive operation was varied after
  each round.

\item[The \textbf{AT\&T library}]~\cite{dgttv-ddage-00} contains 389 undirected and 5114 directed
  real-world graphs with 3--1104 vertices and 1--7602 edges,
  respectively. The latter set contains the North DAGs, which are 1277 acyclic
  connected graphs with 10 to 100 vertices.
  The graphs were collected by Stephen North at the AT\&T Bell Labs by
  running two years an e-mail graph-drawing service.
  The graphs came from very heterogeneous sources,
  mainly representing different phases of various software-engineering
  projects. As a result, the densities of graphs with more or less the same
  number of vertices vary from very sparse to extremely dense, i.e., the
  relative densities are not uniformly distributed over the different
  numbers of vertices of the graphs. When verifying asymptotic running
  times with these graphs it is more appropriate to compare the runtimes with
  the number of edges rather than the number of vertices.
%
  The North DAGs were processed such that for each isomorphism class
  (detected over identical vertex labels) only one representative graph was
  kept. Where necessary, minimal sets of edges were randomly added to make
  the graphs connected. Finally, some edges were heuristically inverted to
  eliminate cycles.

\item[The \textbf{DIMACS challenge graphs}]~\cite{dimacs}
  are a large collection (about 20 GB) of graphs forming the testbed for the
  DIMACS implementation challenges, which started in 1990 and explore
  questions of determining realistic algorithm performances and comparing
  them to theoretical bounds. The addressed problems include graph
  partitioning and clustering, shortest paths, TSP, semidefinite
  optimization, and nearest neighbor searches.
  The instances are real-world graphs (e.g., co-author and citation
  graphs, street networks) and randomly generated graphs (e.g., Delaunay
  graphs, geometric graphs, uniformly drawn Erd\H{o}s-Réni graphs). It
  includes Walshaw's graph partitioning archive~\cite{swc-cesmo-04} and a
  small subset of the Florida sparse-matrix collection.

\item[The \textbf{Florida sparse-matrix collection}]~\cite{dh-ufsmc-11}
  is a large, growing set of sparse matrices that arise
  in real-world applications.
  The collection is widely used for performance analysis by the numerical
  linear-algebra community.
  This set of mostly very large instances originates from a wide spectrum of
  domains, including structural engineering, electromagnetics, semiconductor
  devices, thermodynamics, optimization, circuit simulation, and financial
  modeling.
  The collection currently contains 2541 matrices and the largest matrix
  has a dimension (maximum of the number of rows and columns) of more
  than 100 million.%
  \footnote{See web-search interface at
  \url{http://www2.research.att.com/~yifanhu/GALLERY/GRAPHS/search.html},
  accessed August 2011.}
  The library includes nearly every matrix from the matrix-market
  collection~\cite{bprbd-mmwrt-97}, which additionally includes matrix
  generators.
\end{list}

\paragraph{Evaluation.} An evident weakness of the mentioned libraries is the lack of a
significant number of real-world instances for a wide
variety of applications, as well as the useful contextual information
this would provide, i.e., where the data stems from,
the original use, as well as detailed type or semantic information.
Even though many of the available graphs are---or are derived
from---real-world graphs, they only cover a small set of
applications. Noticeably absent are the important application areas of
biology and social sciences.  Some of the datasets are also quite old,
which means that the graphs do not necessarily represent typical
characteristics or sizes of current real-world data.  A lack of
real-world examples may be attributed to an inability or unwillingness of
practitioners to share their data, a situation which may be improved by
addressing issues of ownership and copyright.
The heavy use of randomly generated graphs is also apparent.
Random graphs play an important role in algorithm
evaluation and engineering, and should therefore also be a part of the
graphbase, possibly along with the corresponding generator code.

\begin{table}[t]\footnotesize
   \centering
  \begin{tabular}{|l|>{\centering}p{1.0cm}|>{\centering}p{1.0cm}|
    >{\centering}p{1.0cm}|>{\centering}p{1.0cm}|>{\centering}p{1.0cm}|>{\centering}p{1.0cm}|}
    \hline
       \backslashbox[12em]{\makebox[12em][l]{\rlap{Services}}}{Systems}
        & \begin{sideways}Tübingen~\cite{ekms-gogds-11}\end{sideways}
        & \begin{sideways}Dortmund~\cite{s-krgev-11}\end{sideways}
        & \begin{sideways}Stanford~\cite{k-sgb-94}\end{sideways}
        & \begin{sideways}Matrix market~\cite{bprbd-mmwrt-97}\mbox{~}\end{sideways}
        & \begin{sideways}DIMACS~\cite{dimacs}\end{sideways}
        & \begin{sideways}Florida~\cite{dh-ufsmc-11} \end{sideways} \tabularnewline
    \hline
    Categorization tags        & Y & Y & Y & Y & Y & Y \tabularnewline
    Further info / comments    & Y & N & Y & N & N & Y \tabularnewline
    Search for tags            & Y & Y & N & Y & N & Y \tabularnewline
    Conversion to file formats & Y & Y$^\ast$ & N & N & N & Y \tabularnewline
    Search for properties      & Y & Y & N$^{\ast\!\ast}$ & N & N & Y \tabularnewline
    Creation method            & N\threestar & N\threestar & Y & N & Y & N \tabularnewline
    Support for images / layout & Y & N & N & Y & N & Y \tabularnewline
    Autom.\ analysis of properties & Y & Y & N & N & N & N \tabularnewline
    Support for web services    & N & N & N & N & N & Y \tabularnewline
    References to publications & Y & Y$^\ast$ & Y & Y & Y & Y \tabularnewline
    Forum                      & N & N\threestar & N & N & N & N \tabularnewline
    Availability of generators & N & N & Y & Y & N & N \tabularnewline
    \hline
    \multicolumn{7}{l}{} \\
    \multicolumn{7}{l}{$^\ast$ implemented but not yet available for the
      user} \\
    \multicolumn{7}{l}{$^{\ast\!\ast}$ search possible only for matrix
      properties} \\
    \multicolumn{7}{l}{\threestar possible as a future extension}
  \end{tabular}\\
  \caption{Functionality of existing systems}
  \label{tab:systems}
\end{table}

\section{A Working Prototype: GraphArchive}
\label{sec:grapharchive}

In this section we give an overview of \emph{GraphArchive},
a platform for exchanging and archiving graphs
meant as a prototype for the Open Graph Archive.
It is developed at the University of Tübingen and was designed as
a successor to \emph{GraphDB}, a now discontinued first attempt at creating a
web-based graphbase.  GraphArchive is an interactive online system built
with modern web technologies.
Below we list the main
features of the existing prototype, followed by a short description of
its software architecture. For more details, we
refer to~\cite{ekms-gogds-11}.  The working system can be accessed
online at
\begin{center}\url{http://graphdrawing.org/grapharchive/}\,.\end{center}



\paragraph{Main features.}
The features of GraphArchive, as listed below, have been chosen to
support the goal of providing an open and easily
accessible system.

\begin{description}
  \item[Web-based user interface.]
  		The user interface is provided via a browser.
		A web portal offers all functionality that is needed to handle
  		graphs, including uploading datasets, inspection and management of
		existing graphs, searching for specific graphs, and downloading datasets.
  		Registration is performed online using a registration
		form, which is processed automatically.  Standard
		techniques are used to prevent registration by spam bots.
  \item[Minimal permission management.]
  		There are no groups of users that define rights for
                small circles of users. Licenses for graphs limiting
                their usage are
  		not encouraged in our open approach. However, if
                necessary, a license can be attached to a selected graph.
  		After confirming registration by going through the opt-in
                e-mail process, a user has access to all graphs and can
  		initiate queries without restrictions.
  \item[Categorization of graphs.]
  		For search queries, graphs can be assigned to the field(s) of application that they originate from. This enables researchers
  		from different fields to use GraphArchive as a common platform.
  \item[Automatic graph analysis.]
  		After upload, graphs (with $<100,000$ vertices) are
		automatically analyzed in order to provide consistent data.
  		Consistency is very important for queries on graph properties.
  \item[Multi-criterion search.]
  		Queries can be performed on multiple parameters, specifying
  		graph properties, categories, author, name, and upload
  		date. Also, parameters can be added later to further
  		narrow down the result set.
  \item[Graph visualization.]
  		An image of a graph is valuable if a user wants to
		visually inspect the properties of a graph. Layouts are
  		computed automatically in the background and can also be changed after upload.
  \item[Unique links to graphs.]  A URI  associated with
    each graph allows for a permanent reference to be used in publications. By
    giving the URI, the user can quickly jump to a particular dataset.
    Reference annotations can be assigned to a graph in order to highlight
    publications and/or websites that refer to or make use of the graph.
  \item[Visual comparison of multiple graphs.]
  		For a quick comparison of graphs, we support simultaneous
		presentation of multiple graphs. Properties
  		are displayed for all graphs. Boolean properties, e.g.,
		directed/undirected, are presented visually on a scale
		(properties can be shared
  		by (a) no graph, (b) a subset of the displayed graphs, or (c) all graphs).
  \item[Several file formats.]
  		When supporting many application domains it is
                impossible to dictate the file
                format used. Therefore, we aim at
  		supporting as many formats as possible. The
		system is extendable and allows for addition of further formats in the future.
  		For downloading graphs, a user can choose the format
                that fits best to his or her work environment. We provide
  		cross conversion between different formats (the users can select any supported
		format and the system performs the conversion automatically).
  \item[Import/export of multiple graphs.]
  		We allow upload/download of several\linebreak{} graphs simultaneously in
        zip-compressed form. In an upload process, each file in a compressed
        archive can be optionally processed individually (for property
  		analysis and layout computation).
  \item[Guest access for non-registered users.]
  		If a user wants to check a specific graph, he or she can access a detailed view of the graph using its URI. All
  		properties and attributes of that graph are made
                visible via a guest account.
 \end{description}

\paragraph{Software architecture.}  GraphArchive is built with
common web technologies.  The application is written in
PHP5\footnote{See project homepage:~\url{http://www.php.net},
  accessed July 2011.} and uses Apache2\footnote{See project
  homepage:~\url{http://www.apache.org}, accessed July 2011.} for
online presentation. For graph analysis and layout computation, we
use the Java graph library yFiles;\footnote{Developed by
  yWorks GmbH: \url{http://www.yworks.com}, accessed July 2011.}
these computations are handled in the background via PHP/Java
bridge.\footnote{See project homepage:
  \url{http://php-java-bridge.sourceforge.net/pjb/index.php},
  accessed July 2011.} Data storage is managed by PostgreSQL
database management system.\footnote{See project homepage:
  \url{http://www.postgresql.org/}, accessed July 2011.}

More details and a descriptive walk-through showing a typical use case of
the system can be found in~\cite{ekms-gogds-11}. For more news and
information on the system and its current development status, please
consult the system website.

\section{Conclusion and Outlook}
\label{sec:conclusions}

We advocated the need for an
open, worldwide graphbase to collect and distribute 
graphs and programs for their generation, analysis, manipulation, and
drawing. Our recommendations for the supported features of such a
system stem from the discussions and experiences within the
graph-drawing community and the results of a survey conducted at
Schloss Dagstuhl.  The specification of reasonable features can
be viewed only as a preliminary wish list---it is expected to change and
grow along with community adoption of the system.  Growth
of the content and evolution of the system will be driven both by the
demands of the users and their willingness to contribute
material. We described a working prototype and propose
that it will be extended and used to build \emph{the} Open Graph Archive.
The prototype already supports many of the recommended features and
fully satisfies the minimum requirements.

Our hope is to stimulate discussion on the initial system
proposal and trigger community growth around the Open Graph Archive.
The success of this project requires a passionate and enthusiastic community.
We urge you to step up and participate by critiquing the existing
system, helping the development effort, or contributing material to
the graphbase.


%




\bibliographystyle{abbrv}
\bibliography{MultiPoster}

\end{document}